\documentclass{amsart}
\usepackage[numbers]{natbib}
\usepackage{amsmath}
\usepackage{amsfonts}
\usepackage{dsfont}
\usepackage{rotating}
\usepackage{adjustbox}  
\usepackage{graphicx}
\usepackage{url}
\usepackage{caption}
\usepackage[ruled, vlined]{algorithm2e}
\usepackage{bbold} 
\usepackage{mathtools}
\usepackage{color}

\usepackage{todonotes}

\DeclareMathOperator*{\argmin}{argmin}

\DeclareMathOperator*{\GPD}{GPD}
\DeclareMathOperator*{\NLL}{NLL}

\numberwithin{equation}{section}

\begin{document}

\title[Spatial FEM-BV-GPD]{On Computationally-Scalable Spatio-Temporal Regression Clustering of Precipitation Threshold Excesses}%

\author{Olga Kaiser}
\address{Universit\`a della Svizerra Italiana, Institute of Computational Science, Via Giuseppe Buffi 13, 6904 Lugano, Switzerland}
\curraddr{NNAISENSE, 6904 Lugano, Switzerland}
\email{olga.kaiser@nnaisense.com}
\thanks{Corresponding author: Olga Kaiser}

\author{Dimitri Igdalov}
\address{Universit\`a della Svizerra Italiana, Institute of Computational Science, Via Giuseppe Buffi 13, 6904 Lugano, Switzerland}
\curraddr{NNAISENSE, 6904 Lugano, Switzerland}
\email{dimitri.igdalov@nnaisense.com}

\author{Olivia Martius}
\address{Institute of Geography, Oeschger Centre for Climate Change Research, University of Bern, Hallerstrasse 12, 3012 Bern,  Switzerland}
\email{olivia.romppainen@giub.unibe.ch}

\author{Illia Horenko}
\address{Universit\`a della Svizerra Italiana, Institute of Computational Science, Via Giuseppe Buffi 13, 6904 Lugano, Switzerland}
\curraddr{}
\email{illia.horenko@usi.ch}
\thanks{}


\begin{abstract}
Focusing on regression based analysis of extremes in a presence of systematically missing covariates, this work presents a data-driven spatio-temporal regression based clustering of threshold excesses. It is shown that in a presence of systematically missing covariates the behavior of threshold excesses becomes nonstationary and nonhomogenous. The presented approach describes this complex behavior by a set of local stationary Generalized Pareto Distribution (GPD) models, where the parameters are expressed as regression models, and a latent spatio-temporal switching process. The spatio-temporal switching process is resolved by the nonparametric Finite Element Methodology for time series analysis with Bounded Variation of the model parameters (FEM-BV). The presented FEM-BV-GPD approach goes beyond strong a priori assumptions made in standard latent class models like Mixture Models and Hidden Markov Models. In addition, it provides a pragmatic description of the underlying dependency structure. The performance of the framework is demonstrated on historical precipitation data for Switzerland and compared with the results obtained by the standard methods on the same data.%
\end{abstract}

\maketitle

%
%
%
\section{Introduction} 
\label{sec:Introduction}
One of the challenges in studying the dynamics of extreme hydrological events like floods and heavy precipitation is the complex behavior of the underlying processes, which act on multiple temporal and spatial scales and exhibit hierarchical organizations. As, in many real application this multiscale behavior can not be fully resolved, the analysis results of such processes can be biased by a presence of a missing (latent) information from unresolved scales. A further challenge is the nature of extreme events: They define the above-average behavior of a process and occur irregularly and rarely. %

To study the behavior of extreme events statistical tools are widely used. One standard statistical tool is provided by the Extreme Value Theory (EVT). EVT is based on the max-stability postulate, which states: The appropriate statistical model for extremes should not only fit the observed data, but also be capable of predicting those extremes that are beyond the observed range~\citep{coles2001introduction}. EVT provides max-stable asymptotic parametric distributions for the intensity of extremes defined as sample maxima and threshold excesses like annual flood levels and heavy precipitation, respectively~\citep{coles2001introduction, Davison2}. The application of EVT assumes that the observations of the process from which the extremes are extracted are independent and identically distributed. The latter assumption implies stationarity of the underlying dynamics. In real applications, this assumptions can not always be met, for example, processes like precipitation depend on the season. The most general way to account for the nonstationarity of the underlying process is to refer to the temporal variability of the parameters of the corresponding distribution for extremes~\citep{davison1990models, coles2001introduction}. %
Further, EVT distinguishes between univariate (marginal) and multivariate modeling. The univariate modeling describes extremes extracted from a single process, for example from a single measurement series obtained from some particular geographical location. The second case refers to extremes extracted from multiple processes and multiple series of extremes that are simultaneously under study~\citep{davison2012statistical, cooley2012survey,huser2014space}. When the different processes correspond to observations at different spatial locations, the multivariate EVT is also known as the spatial EVT. As in this work the applicational part investigates hydrological extremes observed at different locations in Switzerland, we proceed with the notation spatial EVT.%

Based on EVT, univariate Extreme Value Analysis (EVA) can be used to determine the temporal variability of the corresponding distribution parameters. The temporal variability is often associated with the external influences, so called covariates. In this context, the  parameters of the distribution are described by regression models~\citep{davison1990models,coles2001introduction}. One of the standard models is the class of parametric regressions: The model parameters belong to some a priori defined functions depended on covariates, for instance, as linear combinations of a finite set of some explicitly known covariates. To study the nonlinear influence of explicit covariates, standard tools from machine learning, e.g., Support Vector Machines~\citep{Lima2013136} and Artificial Neural Networks (ANNs)~\citep{Cannon}, can be applied. Parametric regression relies on the explicit availability of all of the relevant covariates~\citep{kaiserGPD}. In real applications the complete set of relevant covariates might be not available. Indeed, the selected set of involved covariates is based on the expert knowledge. But, as the driving forces of extreme events are not understood in every detail yet, this selection might neglect relevant covariates that were not associated with the dynamics of extremes so far. %
Further, it is not possible just to involve all the potential covariates that might influence the behavior of extremes: In order to avoid overfitting the number of involved covariates has to be limited. In particular, this applies to regression analysis of extremes because of the relatively small statistics. Thus, focusing on an a priori selected set of available covariates requires an explicit consideration of the eventual impact from the  systematically missing covariates. %

In standard statistical regression analysis the issue of systematically missing information is often addressed as the "unobserved heterogeneity"~\citep{hagenaars2002applied, cameron2013regression}. One standard way to deal with unobserved heterogeneity is to involve the missing covariates into a regression model via a stationary probabilistic error term~\citep{cameron2013regression}. The posterior distribution of the model parameters is then obtained by exploiting Bayesian inference, referring to Bayesian EVA~\citep{macdonald2011flexible, DavisonBoldi}. The results of Bayesian statistics rely strongly on the choice of the prior. Further, the choice of the prior implies assumptions about the systematically missing covariates, for instance, a Gaussian prior implies independent and identically distributed missing covariates. %
Another strategy to address unobserved heterogeneity is to apply latent variable models like finite Mixture Models (MM) and Hidden Markov Models (HMM)~\citep{marcoulides2001new,hagenaars2002applied}. In these models the influence from the missing covariates is reflected by an additional latent variable. Both, MM and HMM are standard techniques in context of EVA, e.g.,~\citep{Betro,DavisonBoldi,macdonald2011flexible}. The final result depends on the a priori assumptions on this latent variable like stationary and Markovian behavior. Additionally, it was shown in~\citep{kaiser2014inference,kaiserGPD} that in the presence of systematically missing covariates parametric approaches will lead to biased results and nonparametric regression techniques are required.%

The class of nonparametric regression includes, for instance, Generalized Additive Models (GAM)\citep{wahba1990spline,hastie1990generalized,Hastie2009Elements}. GAM describe the temporal variability by a linear combination of smooth nonparametric functions of covariates and are widely used to study the behavior of extremes in context of EVA~\citep{yee2007vector,chavez2005generalized,padoan2008mixed}. GAM could be also applied to resolve the nonstationary offset. However, its smoothness might be a restriction, as it implies the locality property and the inability to describe discontinuous functions~\citep{hastie1990generalized,kaiserGPD}. %
An alternative provide the recently introduced FEM-BV-GEV and FEM-BV-GPD approaches~\citep{kaiser2014inference,kaiserGPD}. They are a combination of Finite Element Methodology (FEM) with Bounded Variation (BV) for time series analysis and univariate GEV and GPD. FEM-BV-GEV/FEM-BV-GPD capture the influence of systematically missing covariates by a nonstationary offset process. %

Besides modeling the temporal variability, spatial EVA explores also the dependence structure among extremes which stem from different processes. The dependence structure is measured by the joint probability and reveals how different processes/locations are related and influence each other in terms of the intensity of extremes. The modeling of the spatial dependence structure of the intensity of extremes is an active research field. The state-of-the-art methods, as a combination of EVT and geostatistics, can be roughly classified into max-stable processes and Bayesian Hierarchical Models (BHM)~\citep{davison2012statistical, cooley2012survey,huser2014space}. We discuss very briefly the developments in this field, for a more comprehensive overview please refer to~\citep{davison2019spatial, beirlant2006statistics, davison2012statistical, huser2014space, cooley2012survey,ribatet2013extreme}. %

Max-stable processes provide the spatial extension of the univariate EVT~\citep{de1984spectral,kotz1988multivariate,beirlant2006statistics,huser2014space,bacro2013estimation}. There is no finite parametrization of a max-stable process. That is, the spatial distribution function depends on the a priori choice for the spatial dependence structure. For instance, Smith and Schlather max-stable processes assume Gaussian behavior of the the underlying spatial process~\citep{schlather2002models, kabluchko2009stationary}. In order to model anisotropic and/or nonstationary dependence, the Gaussian process incorporates parametric spatio-temporal covariance functions~\citep{davison2012statistical, huser2014space}. %

BHM describe a complex system by decomposing its dynamics into different layers of parameter variation~\citep{LIMA2010307, cressie2011statistics}, where a priori parametric assumptions about each layer are required. Standard BHM approaches are based on conditional independence and thus may not capture the underlying spatial dependence structure~\citep{davison2012statistical, cooley2012survey}. In order to account for the dependence structure, more sophisticated formulations were introduced, based, e.g., on Gaussian copula models~\citep{sang2010continuous}. Other BHM based, e.g., on the skew-t processes~\citep{morris2017space} or on spatial Markov models~\citep{reich2018spatial} account for a local asymptotical dependence and long-distance asymptotic independence by a spatial partitioning/clustering of the considered spatial domain. %
Nonparametric BHM extensions belong to the class of mixture models, e.g., the max-stable hierarchical model~\citep{reich2012hierarchical}, Dirichlet-based copula~\citep{fuentes2012nonparametric}. Both techniques approximate the dependence structure by a linear combination of a priori defined kernel functions and make a priori parametric assumptions on the distribution of the approximation coefficients~\citep{reich2012hierarchical}. %
 
The above discussed approaches enable not only the analysis of the underlying spatial dependence structure but also provide the possibility to make inference about extremes on not gauged locations. However, in one way or another, these approaches dwell on a priori assumptions about the underlying spatial dependence structure, either directly like in the case of max-stable processes or implicitly by choosing the appropriate kernel function as in the case of nonparametric BHM. These a priori assumptions also imply a priori assumptions about the systematically missing covariates, which might eventually lead to a bias in real applications. %

In this paper we would like to contribute towards data-driven spatio-temporal analysis of extreme events. First, we address explicitly the challenge of systematically missing covariates in the context of spatio-temporal regression analysis of threshold excesses and present a solution that goes beyond standard a priori assumptions made for instance in MM and HMM based techniques  by assuming that the switching process is from the space of functions with bounded variation. Second, we present a stable and robust Matlab framework\footnote{Matlab code: \url{https://github.com/vonera/FEM_GPD}} that provides an optimal sparse solution for the spatio-temporal clustering problem of threshold excesses, by incorporating information theory based model selection criteria and L1-regularized regression. %

The paper is organized as follows. Section~\ref{sec:methodology} presents the FEM-BV-GPD methodology. Section \ref{sec:comp_framework} focuses on the corresponding computational framework. Section~\ref{sec:conceptual_comparison_to_standard_methods} discusses the advantages and limitations of the FEM-BV-GPD framework. Section~\ref{sec:Real_Application} demonstrates the performance of the proposed framework on meteorological data - on historical extreme precipitation measurement series over Switzerland. Section~\ref{sec:conclusion} contains the conclusion and the outlook.%
%
%
%
%
%
%
%
\section{Methodology} 
\label{sec:methodology}
Consider a series of observations of a process like precipitation or temperature measured at different locations for a fixed period of time. Then, for each location the threshold excesses are extracted with respect to a fixed threshold, e.g., to the $0.98$ quantile. The resulting series of threshold excesses is denoted by $Y_{s,t}$, with $Y_{s,t}>0$; here the index $s$ stands for different locations, with $s=s_1,\dots,S$, and $t$ represents the time steps when the excesses were observed, with $t=t_1,\dots,t_{T_s}$. Please note, by definition, the length of $Y_{s,t}$, i.e., $t_{T_s}$, might vary from location to location. Further, we assume, that the marginal distribution of $Y_{s,t}$, for each $s$, is the fully nonstationary GPD, described by its pdf:%
\begin{align}\label{eq:GPDpdf} 	
	h\left(Y_{s,t};\xi\left( s,t \right),\sigma\left(s,t\right)\right) =
	\begin{cases} 	
		 \frac{1}{\sigma\left(s,t\right)}\left(1 + \frac{\xi\left(s,t\right) Y_{s,t}}{\sigma\left(s,t\right)}\right)^{-\frac{1}{\xi\left(s,t\right)}-1} &,\, \xi(s,t)\neq 0,\\ \frac{1}{\sigma\left(s,t\right)}\exp{\left(-\frac{Y_{s,t}}{\sigma(s,t)}\right)} &,\, \xi(s,t) = 0,%
	\end{cases}
\end{align}
where $\xi(s,t)$ is the shape and $\sigma(s,t)$ the scale parameter with constraints %
\begin{align}\label{eq:constrParamGPD}
		\left[1+\frac{\xi(s,t) Y_{s,t}}{\sigma(s,t)}\right]>0,\quad \sigma(s,t)>0\quad \text{and}\quad \xi(s,t)\in(-0.5,0.5)\quad \forall s,t.%
\end{align}
In the following, the GPD model parameters are summarized by $\Theta(s,t) = \left(\xi(s,t),\sigma(s,t)\right)$. Then, $\Theta(s,t)$ is estimated by deploying the likelihood estimator. The constraint $\xi(s,t)\in(-0.5,0.5)$ is required, in order to insure the "regularity" of the likelihood estimator, i.e., Gaussian distribution of the estimator with true model parameters as the mean and with the Fischer information matrix as the variance~\citep{coles2001introduction, davison2003statistical}.%

In the next step we parametrize $\Theta(s,t)$, by assuming that $\Theta(s,t)$ is a spatio-temporal process which addresses both the marginal temporal variability of $Y_{s,t}$ and its spatial dependence. Thereby, we investigate the idea that this process is mainly governed by covariates. That is, there exists a set of principal covariates that controls both the marginal temporal variability and the spatial dependence structure of the threshold excesses. %
While regression models are widely used to describe the temporal marginal behavior of extremes, the idea to describe the spatial dependence by covariates is underpinned by the fact, that weather systems such as cyclones and anticyclones influence not just single locations but rather entire regions. For instance, spatially-persistent blocking anticyclones can be responsible for heat waves over extended areas, e.g., ~\citep{feudale2011influence,cassou2005tropical}. Other studies associate regional heavy precipitation with large-scale atmospheric flow patterns~\citep{wang2005observed,jones2014sensitivity,plaut2001heavy}.%

Let $U^{all}_{s,t} = \left(u^{all}_1(s,t),\dots,u^{all}_{\mathcal J}(s,t)\right)$ be the set of all the principal covariates. $U^{all}_{s,t}$ is composed of (a) local covariates observed at each location, responsible for local temporal changes, e.g., temperature, and (b) global covariates being the same for all locations, influencing, among others, the overall spatial dependence structure. Then, we assume that conditioned on $U^{all}_{s,t}$, provided that all $U^{all}_{s,t}$ are known and observed, the process $Y_{s,t}$ becomes independent in space and time, i.e.,%
\begin{align}\label{eq:allcovaYinde}
	Y_{s,t}|U^{all}_{s,t}\overset{independent}{\sim}\GPD\left(Y_{s,t};\Theta\left(s,U^{all}_{s,t}\right)\right),%
\end{align}
the pdf of $\GPD$ and the model parameters $\Theta\left(s,U^{all}_{s,t}\right)$ are defined in equation (\ref{eq:GPDpdf}). The limitation of the "conditional independence" assumption is discussed in Section~\ref{sec:conceptual_comparison_to_standard_methods}. The optimal $\Theta\left(s,U^{all}_{s,t}\right)$ can be obtained through constrained minimization of the corresponding Negative Log-Likelihood function (NLL). Since $Y_{s,t}$ is conditionally independent for given $U^{all}_{s,t}$, the corresponding likelihood is the product, and NLL is the sum, of the negative marginal likelihoods over all locations and time steps: %
\begin{align}
	\label{eq:YNLL}
	\NLL\left(Y_{s,t}; \Theta\left(s,U^{all}_{s,t}\right)\right) = -\sum\limits_{i=1}^{S_{\vphantom{s_i}}}\sum\limits_{j=1}^{T_{s_i}} \log\left(h\left(Y_{s_i,t_j};\Theta\left(s_i,U^{all}_{s_i,t_j}\right)\right)\right).%
\end{align}

In many real applications the principal set of covariates $U^{all}_{s_i,t_j}$ in equation (\ref{eq:allcovaYinde}) is not known. Thus, the spatio-temporal behavior of $Y_{s,t}$, i.e., $\Theta\left(s,U^{all}_{s,t}\right)$, needs to be parametrized in presence of systematically missing covariates. For this purpose we extend the univariate FEM-BV-GPD approach with nonstationary regression of threshold excesses in a presence of systematically missing covariates~\citep{kaiserGPD} towards spatial FEM-BV-GPD.%

Following the univariate FEM-BV-GPD, the model parameters are parametrized by a linear combination of $U^{all}_{s,t}$. Linear regression is preferred, as it allows direct interpretation of the influence of the covariates on the behavior of $\Theta\left(s,U^{all}_{s,t}\right)$ and can be extended towards nonlinear regression by incorporating nonlinear covariates. We divide the set of principal covariates into observed ones, denoted by $U_{s,t}=\left(u_1(s,t),\dots, u_P(s,t)\right)$ and systematically missing ones, denoted by $U^{miss}_{s,t}=\left(u^{miss}_1(s,t),\dots, u^{miss}_Q(s,t)\right)$. Such that, the linear regression model, here exemplified on the shape parameter, is written down as%
\begin{align}
	\label{eq:ksi_full}
	\begin{split}
		\xi(s,U^{all}_{s,t}) = \xi_0(s) + \sum\limits_{p=1}^P \xi_p(s)u_p(s,t) + \frac{1}{Q}\sum\limits_{q=1}^Q \xi_q(s)u^{miss}_q(s,t).%
	\end{split}
\end{align}
The normalization/scaling of the missing covariates entails no loss of consistency - it will be absorbed by the coefficients. In order to get rid of $U^{miss}_{s,t}$, the equation (\ref{eq:ksi_full}) is completed with the averaged behavior of systematically missing covariates $\mathbb E[u^{miss}_q(s,t)]$, such that we get: %
\begin{align}
	\label{eq:ksi_full_e}
	\begin{split}
		\xi(s,U^{all}_{s,t}) &= \xi_0(s) + \sum\limits_{p=1}^P \xi_p(s)u_p(s,t)\\ &+\frac{1}{Q}\sum\limits_{q=1}^Q \xi_q(s)\left(u^{miss}_q(s,t) - \mathbb E[u^{miss}_q(s,t)]\right) + \frac{1}{Q}\sum\limits_{q=1}^Q \xi_q(s) \mathbb E[u^{miss}_q(s,t)].%
	\end{split}
\end{align}
The third sum in equation (\ref{eq:ksi_full_e}) can be reduced to a Normal distributed error term: Under the assumption that $u_q^{miss}(s,t)$ are independent for all time steps and fulfill the Lindeberg condition, the Central Limit Theorem for independent variables can be applied as $Q\rightarrow\infty$. Further, the first and the last terms in equation (\ref{eq:ksi_full_e}) are combined to %
\begin{align}
	\label{eq:nonstat_nonhom_offset}
	\xi_0(s,t) = \xi_0(s) + \frac{1}{Q}\sum\limits_{q=1}^Q \xi_q(s) \mathbb E[u^{miss}_q(s,t)],
\end{align}
denoted in the following as the offset. Finally, the reduced stochastic regression model with a nonstationary, nonhomogenous offset is given by%
\begin{align}
	\label{eq:ksiNonstatNonhom}
	\xi(s,t,U_{s,t}) = \xi_0(s,t) + \sum\limits_{p=1}^P \xi_p(s)u_p(s,t) +\epsilon(s,t), \quad \epsilon(s,t)\sim\mathcal N(0,\tau(s,t)).%
\end{align}
The reduced regression model for the scale parameter is obtained exactly analogue: %
\begin{align}
	\label{eq:sigmaNonstatNonhom}
	\sigma(s,t,U_{s,t}) = \sigma_0(s,t) + \sum\limits_{p=1}^P \sigma_p(s)u_p(s,t) +\hat\epsilon(s,t), \quad \hat\epsilon(s,t)\sim\mathcal N(0,\hat\tau(s,t)).%
\end{align}
In the case when the covariates are not independent, the Karhunene-Lo\`eve transformation can be applied for an orthogonal representation of the covariates, i.e., for their decorrelation~\citep{Kozek,loeve1978probability}.%

The parameterization in equations (\ref{eq:ksiNonstatNonhom}) and (\ref{eq:sigmaNonstatNonhom}) reflects the influences coming from unobserved covariates by the spatio-temporal offsets $\sigma_0(s,t)$ and $\xi_0(s,t)$ and the corresponding noise terms, $\epsilon_{s,t}\sim\mathcal N\left(0,\tau_{s,t}\right)$ and $\hat\epsilon_{s,t}\sim\mathcal N\left(0,\hat\tau_{s,t}\right)$, respectively. The noise terms imply a Gaussian prior for the model parameters with no closed posterior for the parameters. For simplification, in line with the univariate FEM-BV-GPD, the Gaussian noise terms $\epsilon_{s,t}$ and $\hat\epsilon_{s,t}$ in equations (\ref{eq:ksiNonstatNonhom}-\ref{eq:sigmaNonstatNonhom}) are neglected. Motivated among others by the Bernstein-von Mises Theorem: It states that if "enough" data is available, then, regardless of the prior, the posterior distribution of the parameters is asymptotically Gaussian. Thereby, the true parameters act as the mean and the corresponding inverse Fischer information matrix as the covariance matrix~\citep{van2000asymptotic}. Neglecting the noise allows a deterministic parameterization of the model parameters, summarized by %
\begin{align}\label{eq:ThetaNonStatNonHom_gev}
	\Theta(s,t,U_{s,t})=\left(\xi_0(s,t),\dots,\xi_P(s,t),\sigma_0(s,t),\dots,\sigma_P(s,t)\right).%
\end{align}
Please note, in equation (\ref{eq:ThetaNonStatNonHom_gev}) also the regression coefficients are time dependent (without loss of generality). This enables to account for a possibly changing influence of the involved covariates and is validated especially when the process $Y_{s,t}$ was observed over a long period of time. %

In the next step the regression coefficients are described by exploiting nonparametric techniques. Thereby we focus on the Finite Element time series analysis methodology (FEM)~\citep{Horenko,Metzner}. In particular, the idea proposed in~\citep{deWiljes2} is adapted: The spatio-temporal parameter dynamics is interpolated by a set of $K\geq 1$ locally stationary parameters and a persistent binary spatio-temporal switching process $\Gamma(s,t)=\left(\gamma_1(s,t),\dots,\gamma_K(s,t)\right)^{\dagger}$ with $\gamma_k(s,t)\in\lbrace 0,1\rbrace$, for $k=1,\dots,K$. The persistency assumption is motivated by the observation that many realistic problems demonstrate a persistent (metastable or regime-switching) behavior of their parameters. Further, for avoiding probabilistic a priori assumptions on $\Gamma(s,t)$, e.g., stationarity or Markovian behavior, $\Gamma(s,t)$ is assumed to be in the space of function with Bounded Variation (BV) with respect to time and is discretized by Finite Elements in line with the FEM approach~\citep{Horenko,Metzner}. That is, the application of spatial FEM for the parametrization of the regression coefficients in (\ref{eq:ksiNonstatNonhom}) and (\ref{eq:sigmaNonstatNonhom}) results in%
\begin{align}
	&\xi(s,t,U_{s,t}) \approx \sum\limits_{k=1}^K\gamma_k(s,t)\xi_k(U_{s,t}), \text{ with }\xi_k(U_{s,t}) = \xi_{k0} + \sum\limits_{p=1}^P \xi_{kp}u_p(s,t),\label{eq:sigma_nonstat}\\%
	&\sigma(s,t,U_{s,t}) \approx \sum\limits_{k=1}^K\gamma_k(s,t)\sigma_k(U_{s,t}), \text{ with }\sigma_k(U_{s,t}) = \sigma_{k0} + \sum\limits_{p=1}^P \sigma_{kp}u_p(s,t)\label{eq:ksi_nonstat},%
\end{align}
with constraints on $\xi_k(U_{s,t})$ and $\sigma_k(U_{s,t})$ in line with constraints in (\ref{eq:constrParamGPD}). The persistency constraint is insured by%
\begin{align}\label{eq:ConstraintGammaTemp_BV}
	 \|\gamma_k(s,.)\|_{BV([t_1,t_{T_s}])} = \sum\limits_{j=1}^{T_s-1} |\gamma_k(s,t_j+1) - \gamma_k(s,t_j)| \leq C_k(T_s),\,\forall s,k. %
\end{align}
where for a fixed $s$, $C_k(T_s)$ denotes the maximal number of allowed transitions between the model/cluster $k$ and all the other models/clusters in the time interval $[t_1,t_{T_s}]$. Further on we will refer to $C(T) = \max\limits_{s,k}{C_k(T_s)}$. The natural boundary of $C(T)$ is given by the maximal value of $T_s$ over all locations. Also spatial regularization of the switching process would make sense, but for computational reasons this work focuses on temporal persistency only. Application and adaptation of a computationally-feasible spatial regularization in the FEM-BV-GPD context remains for future work. %

In summary, the parameterization of the spatio-temporal behavior of $Y_{s,t}$ is given by parameters $\Theta$ and $\Gamma(s,t)$, with $\Theta = \left(\theta_1,\dots,\theta_K\right)$ where $\theta_k = \left(\xi_{k0},\dots,\xi_{kP},\sigma_{k0},\dots,\sigma_{kP}\right)$. Please note, in equation (\ref{eq:allcovaYinde}) the assumption was made that $Y_{s,t}$ are conditionally independent given the set of all the principle covariates $U_{s,t}^{all}$. Now, as the spatio-temporal switching process $\Gamma(s,t)$ reflects the systematically missing covariates, given the observed set of covariates $U_{s,t}$ and the appropriate $\Gamma(s,t)$, the threshold excesses $Y_{s,t}$ are conditionally independent for all time steps $t$ and locations $s$  , i.e.:%
\begin{align}\label{eq:GammaCovaYinde}
	Y_{s,t}|U_{s,t},\Gamma(s,t)\overset{independent}{\sim}\GPD\left(Y_{s,t};\left(\Theta,\Gamma(s,t)\right)\right).%
\end{align}
The optimal parameter set $\left(\Theta,\Gamma(s,t)\right)$ is obtained by constrained minimization of the NLL: %
\begin{align}\label{eq:YNLL_gamma}
	\NLL\left(Y_{s,t}; \Theta,\Gamma(s,t)\right) = -\sum\limits_{i=1}^{S_{\vphantom{s_i}}}\sum\limits_{j=1}^{T_{s_i}} \log\left(h\left(Y_{s_i,t_j};\Theta,\Gamma(s_i,t_j)\right)\right),%
\end{align}
with respect to persistency constraints on $\Gamma(s,t)$ and corresponding constraints on $\Theta$. As the switching process is restricted to binary values, i.e., $\Gamma(s,t)\in\lbrace 0, 1\rbrace$, it can be carried outside the model parameters, such that: %
\begin{align}\label{eq:YACF_gamma_gpd}
	\NLL\left(Y_{s,t}; \Theta,\Gamma(s,t)\right) = -\sum\limits_{i=1}^{S_{\vphantom{s_i}}}\sum\limits_{j=1}^{T_{s_i}}\sum\limits_{k=1}^{K_{\vphantom{s_i}}} \gamma_k(s_i,t_j)\log\left(h\left(Y_{s_i,t_j};\theta_k\right)\right).%
\end{align}
In this way, the model parameters $\Theta$ and $\Gamma(s,t)$ become uncoupled and NLL can be minimized by alternating optimization (AO) and standard optimization techniques. More details can be found in Section~\ref{sec:comp_framework}. Please note, in the general FEM approach $\Gamma(s,t)$ is defined in the range $[0,1]$. However, in this case $\Gamma(s,t)$ remains in the model parameters and the minimization of the corresponding NLL is difficult. In this work the computation efficiency dominates the decision for a binary $\Gamma(s,t)$. This decision implies that the resulting model becomes a mixture model where the latent variable, i.e., $\Gamma(s,t)$, is obtained without strong a priori assumptions. %

Before writing down the final spatial FEM-BV-GPD formulation, an additional regularization constraint is added on $\Theta$: In order to identify the most significant subset of available covariates, FEM-BV-GPD incorporates Lasso shrinkage on the GPD model parameters. By constraining the $L1$ norm of $\Theta$ the coefficients of insignificant covariates are forced towards zero values. The Lasso regularization is incorporated via a Lagrange multiplier $\lambda$. The final spatial FEM-BV-GPD approach can be formulated as the minimization of %
\begin{align}
	NLL_{\lambda}(Y_{s,t}; \Theta,\Gamma(s,t))= -\sum\limits_{i=1}^{S_{\vphantom{s_i}}}\sum\limits_{j=1}^{T_{s_i}}\sum\limits_{k=1}^{K_{\vphantom{s_i}}} \gamma_k(s_i,t_j)\log\left(h\left(Y_{s_i,t_j};\theta_k\right)\right) + \lambda \|\Theta\|_{L1},\label{eq:YACF_gamma_gpd_lambda}%
\end{align}	
with respect to constraint on $\Gamma(s,t)$:	
\begin{align}	
	\gamma_k(s,t)\in\lbrace 0,1\rbrace\quad\forall s,k,t,\quad \|\gamma_k(s,t)\|_{BV([t_1,t_{T_s}])} \leq C(T),\quad\forall s,k,\label{eq:constraint_spatGPDI} %
\end{align}
and with respect to constraints on $\Theta$: %
\begin{align}\label{eq:constraint_spatGPDII}
			\left[1+\frac{\xi_k(U_{s,t}) Y_{s,t}}{\sigma_k(U_{s,t})}\right]>0,\quad \sigma_k(U_{s,t})>0\quad\text{and}\quad\xi_k(U_{s,t})\in(-0.5,0.5)\quad%
			 \forall \, U_{s,t}, k. %
\end{align}
Concluding, the application of the spatial FEM-BV-GPD approach to a spatio-temporal series of threshold excesses results in a set of $K\geq 1$ locally stationary model parameters $\left(\theta_1,\dots,\theta_K\right)$ and a spatio-temporal switching process $\Gamma(s,t)$. And while the model parameters are accessible for all locations, the nonstationary switching process is assigned to each location separately (each location $s$ is associated with a $\Gamma(s,t)$, which describes the temporal affiliation to one of the models).%
%
%
%
%
%
\section{Computational Framework} 
\label{sec:comp_framework}
%
This section discusses the computational framework (available on email request) for the constraint minimization of the functional $\NLL_{\lambda}\left(Y_{s,t}; \Theta,\Gamma(s,t)\right)$ derived in (\ref{eq:YACF_gamma_gpd_lambda}). For fixed $K,C(T),\lambda$ the functional (\ref{eq:YACF_gamma_gpd_lambda}) is not convex and there exists no analytical global solution of the constrained minimization problem (\ref{eq:YACF_gamma_gpd_lambda}-\ref{eq:constraint_spatGPDI}-\ref{eq:constraint_spatGPDII}). Instead, a local optimal solution can be found through AO with respect to $\Gamma(s,t)$ and $\Theta$, please compare Algorithm~\ref{algo:FEM-BV-annealing}, lines \ref{lines:subspace1} to \ref{lines:subspace3}. %
%
\IncMargin{1em}
\begin{algorithm}[ht]
\DontPrintSemicolon
\SetKwData{Left}{left}
\SetKwData{This}{this}
\SetKwData{Up}{up}
\SetKwFunction{Union}{Union}
\SetKwFunction{FindCompress}{FindCompress}
\SetKwInOut{Input}{input}
\SetKwInOut{Output}{output}
\caption{getOptimalParameterSet(); Restarting and Alternating Optimization}
\label{algo:FEM-BV-annealing}
\Input{$Y_{s,t}$, $U_{s,t}$, $\lbrace K,\,C(T),\,\lambda\rbrace$, AO convergency threshold value: $Tol$, maximal number of restarts: $maxRestart$, maximal number of AO iterations: $maxAO$ }%
\Output{Global optimal parameter set $\left(\Theta^*,\Gamma^*(s,t)\right)$}
\BlankLine
\lnl{1}$\NLL_{\lambda}\left(Y_{s,t}; \Theta,\Gamma(s,t)\right) = inf$\;
\lnl{2Init}\For{r = 1:maxRestart}{
		\BlankLine
		\lnl{gold}$\Gamma_{opt}(s,t)$; Random generation with respect to persistency constraints\;%
		\lnl{told}$\Theta_{opt} = \argmin\limits_{\Theta}\mathcal L_{\lambda}\left(\Theta ,\Gamma_{opt}(s,t)\right)$\;%
		\BlankLine
		\lnl{lines:subspace1}\While{not convergency or maxAO}%
			{
				\BlankLine
				\lnl{lines:subspace2}\textbf{Step1}:			
				$ \Gamma_{opt}(s,t) = \argmin\limits_{\Gamma(s,t)} \NLL_{\lambda}\left(Y_{s,t}; \Theta_{old},\Gamma(s,t)\right)$; The constrained minimization wrt. $\Gamma(s,t)$ results for BV-regularization in a linear problem, standard methods, e.g., simplex method, can be applied. \;	%
				\BlankLine
				\lnl{lines:subspace3}\textbf{Step2}: 
				$\Theta_{opt} = \argmin\limits_{\Theta} \NLL_{\lambda}\left(Y_{s,t}; \Theta,\Gamma_{opt}(s,t)\right)$; The minimization with Lasso regularization wrt. $\Theta$ is carried out by applying MCMC method, a detailed description is given in~\citep{kaiser2014inference}.%
			}
		\BlankLine
		\lnl{ifsubspace}\If{$ \NLL_{\lambda}\left(Y_{s,t}; \Theta^*,\Gamma^*(s,t)\right) >\NLL_{\lambda}\left(Y_{s,t}; \Theta_{opt},\Gamma_{opt}(s,t)\right)$}%
			{
				\lnl{t*}$\Theta^* = \Theta_{opt}$\;
				\lnl{g*}$\Gamma^*(s,t) = \Gamma_{opt}(s,t)$\;
			}
		}
\BlankLine
\end{algorithm}
\DecMargin{1em}
%
In the following the two involved alternating steps are outlined briefly.

In the first AO step, after the initialization step, the optimal switching process $\Gamma(s,t)$ is estimated: In line with the general FEM approach~\citep{Metzner}, $\Gamma(s,t)$ is discretized by Finite Elements as proposed in~\citep{Horenko}. The $BV$ constraint is involved by adding slack variables, such that a linear constrained minimization problem with respect to $\Gamma(s,t)$ for every $s$ is obtained. Standard algorithms like the simplex method can be applied. Here, the solver available in the Gurobi Toolbox is used~\citep{gurobi}. %

The second AO step estimates the optimal model parameter $\Theta$ for fixed $\Gamma(s,t)$. $\Theta$ is obtained through a gradient-free optimization technique based on the Markov Chain Monte Carlo (MCMC) method. A gradient-free approach performs better in the situations when the derivatives of the functional $\NLL_{\lambda}\left(Y_{s,t}; \Theta,\Gamma(s,t)\right)$ in (\ref{eq:YACF_gamma_gpd_lambda}) with respect to $\Theta$ become very large or unbounded. %

The convergency of the AO is achieved if the value $\NLL_{\lambda}\left(Y_{s,t}; \Theta,\Gamma(s,t)\right)$ stops decreasing significantly. The result of one AO is a local optimal solution, which is dependent on the initialization of the start parameters. In order to obtain the global optimal solution, the whole solution space needs to be explored as far as possible. Deterministic exploration becomes computationally infeasible with increasing dimension of the problem, for instance, when the number of observed locations is increasing. Thus, random exploration is required~\citep{Tarantola}. In the general FEM framework, this random exploration is implemented by starting the local optimization several times with random initializations of the switching process (assuming that the chosen number of trials is high enough to provide the global optimal solution). %

The FEM-BV-GPD framework is scalable and computationally efficient, as both steps in the above AO can be parallelized easily. First, the switching process is uncoupled in the space dimension and thus can be estimated for each location separately in every iteration of AO. Second, the  MCMC-based optimization can be started simultaneously with random starting values for a more wider and faster exploration of the parameter space. %

Finally, the above described AO, in combination with random exploration of the solution space, is applied to different combinations of $K,C(T),\lambda$. Such that in total a set of different optimal models is obtained. The optimal one, i.e., the optimal combination of $K,C(T),\lambda$, is chosen with respect to Information Criteria, for instance Akaike Information Criteria. Here the focus lies on the second order Akaike Information Criteria (AICc). AICc penalizes the short sample size (necessary when dealing with univariate analysis of extremes) and converges to AIC with an increasing sample size~\citep{Kenneth}. Please note, it was recently shown that in Bayesian context the standard model selection criteria, like AIC and Bayesian Information Criteria (BIC) might lead to biased results~\citep{schoniger2014model}. This is also true if the prior information is not available, as the AIC/BIC do not account for the correlation in the model parameters when estimating the number of involved model parameters~\citep{schoniger2014model}. However, as there is no reliable alternative, we focus on AIC. For model validation the Quantile-Quantile (QQ) plot is used~\citep{coles2001introduction}.  %

%
%
%
%
\section{Discussion} 
\label{sec:conceptual_comparison_to_standard_methods}
The proposed FEM-BV-GPD approach is a data driven spatio-temporal regression based clustering of threshold excesses based on resolved covariates only. The nonstationarity and the nonhomogeneity are resolved by incorporating the spatial FEM-BV formulation, i.e., describing the underlying behavior by a set of $K\geq 1$ locally stationary models and a spatial nonstationary switching process $\Gamma(s,t)$, where $K > 1$ indicates the existence of systematically missing covariates. The associated FEM-BV-GPD NLL in (\ref{eq:YACF_gamma_gpd}) provides a locally stationary approximation of the true NLL in (\ref{eq:YNLL}). %
Being a part of the FEM-BV model family FEM-BV-GPD goes beyond strong a priori probabilistic and deterministic assumptions typical for standard approaches. For instance, in contrast to Hidden Markov Models (HMM) and Gaussian Mixture Models (GMM) no Markovian nor Gaussian assumptions about the switching process are made a priori~\citep{Metzner}. As opposite, for instance, to Markov based clustering models, e.g., the spatial Markov model for extremes~\citep{reich2018spatial} and the BHM based on the skew-t processes~\citep{morris2017space} deploy the random partition model for the identification of spatial clusters of extreme events, resulting in a Markovian model for the clustering. %

The switching process $\Gamma(s,t)$ is a nonparametric, nonstationary, bounded process. The natural boundary of $C(T)$ for each location is given by $T_s$ (the number of threshold excesses). Thus, the boundary constraint on $\Gamma(s,t)$ does not confine the solution space of the final optimization problem. The discontinuous behavior of $\Gamma(s,t)$ is not considered as a strong limitation, as in many real climatological processes the underlying dynamics exhibit regime behavior~\citep{franzke2011persistent}. %
Further, FEM-BV-GPD corresponds to adaptive multimodal optimization, as the switching process $\Gamma(s,t)$ allows to consider all observations that exhibit similar behavior as one data set. That is, the local GPD model parameters for each cluster are fitted to data from all locations which are assigned to this particular cluster. This naturally increases the sample size and thus the quality of the ML estimator. This becomes in particular important in the context of the "Battle of extreme value distributions", where it was shown that in real applications the asymptotical results of EVT underestimate the shape parameter of the corresponding distribution due to the limited sample size~\citep{WRCR012557,WRCR20707}. %
The involved local stationary regression identifies the most significant covariates that influence/precede the behavior of extremes. Such that, FEM-BV-GPD can be employed as a robust exploratory regression analysis tool for spatio-temporal extremes. Further, the FEM-BV-GPD is able to account for some nonlinear behavior by the sequence of piecewise-linear approximations in $K$ local regimes. However, the linearity assumption in the regression problem formulation may become a weakness as soon as the influence of covariates is strongly nonlinear. One way to address this, is to involve nonlinear terms, which describe the interacting couplings of the covariates (e.g., $u_1^2(t), u_1(t)\cdot u_2(t),\dots$), as additional covariates. %

A conceptual limitation of the resulting FEM-BV-GPD framework is its underlying assumption about the conditional independence of variables. This conditional independence, also known as local independence, is the underlying assumption in the widely used mixture and latent class models~\citep{hagenaars2002applied}. These models explain the relationship/dependence between the variables by the latent variable, i.e., in context of FEM-BV-GPD by the latent the switching process $\Gamma(s,t)$. Such that this assumption is reduced to the assumption that the variables are independent within a cluster, which is in line with state-of-art methods in EVA like Hidden Markov and Gaussian Mixture based models, e.g.,~\citep{Betro,DavisonBoldi,macdonald2011flexible}. Moreover, even if the temporal dependence within a cluster is ignored, the ML estimator remains unbiased with underestimated standard errors~\citep{fawcett2007improved}. %
Despite the independence assumption, the resulting switching process provides a posteriori insight into the spatial dependence structure: FEM-BV-GPD provides a pragmatic description of the corresponding spatial dependence structure by grouping together all locations that exhibit similar behavior in $\Gamma(s,t)$, e.g., by estimating the Event Synchronization (ES) measure matrix~\citep{malik2012analysis} which allows to identify contiguous regions that exhibit similar spatio-temporal behavior. %
The identification of spatial contiguous regions was also addressed for instance in~\citep{sun2015global}: The proposed methodology was applied by the authors to classify the annual maximum daily precipitation in the U.S. into seven geographically contiguous regions, where the optimal number of clusters was obtained by deploying an information criteria approach. %

A further limitation of FEM-BV-GPD, is that there is no model which can be used for a detailed investigation of the spatial dependency: neither it is possible to measure the strength of the spatial dependence, nor can the resulting description be used for spatial interpolation of missing locations without additional analysis of the resulting switching process. In contrast, these issues can be directly approached by the parametric/nonparametric max-stable models~\citep{davison2012statistical, cooley2012survey,reich2012hierarchical}. The appropriate analysis of the FEM-BV-GPD switching process remains for future work by exploiting, for instance, the results obtained in the field of complex causality networks~\citep{malik2012analysis,causalityGerberHorenko}. %
%
%
%
%
%
%
\section{Application} 
\label{sec:Real_Application}
In this section the FEM-BV-GPD framework is applied to precipitation data. We skip here the demonstration of the robustness of the spatial FEM-BV-GPD approach with respect to systematically missing covariates on a test case. Instead we refer to~\citep{kaiserGPD}, where the univariate FEM-BV-GPD approach was presented. There, the approach was compared to a nonparametric method, based on generalized additive models. The results, regarding the performance in the presence of systematically missing covariates, are transferrable to the spatial FEM-BV-GPD: Nonparametric generalized additive models resolve the influence coming from the systematically missing covariates, i.e., the nonstationary offset, by the smoothing spline. However, the involved smoothing spline does not distinguish between the nonstationarity induced by observed and missing covariates, and has difficulties if the underlying dynamics exhibits jump behavior. %

\begin{figure}[ht]
  \begin{adjustbox}{addcode={\begin{minipage}{\width}}{\caption{%
      Panel (a) shows the swiss map of the 17 weather stations. Shading indicates the height above sea level. The full names of the locations are: G\"uttingen (GUT), Basel/Binningen (BAS), Z\"urich/Kloten (KLO), Fahy (FAH), Glarus (GLA), Napf (NAP), Neuchatel (NEU), Bern (BER), Altdorf (ALT), Chur (CHU), Davos (DAV), Scuol (SCU), Ulrichen (ULR), Nyon/Changins (CGI), Geneve-Cointrin (GVE), Sion (SIO), Lugano (LUG). Panels (b), (c) and (d) display the stationary ES and the ES for the first and the second cluster, respectively. The ES values are rounded to two decimal points. For a better visualization, ES values smaller than 0.25 and bigger than 0.5 are printed in white.}\label{fig:ES_cl12}%
  		\end{minipage}},rotate=90,center}
      \includegraphics[scale=.22]{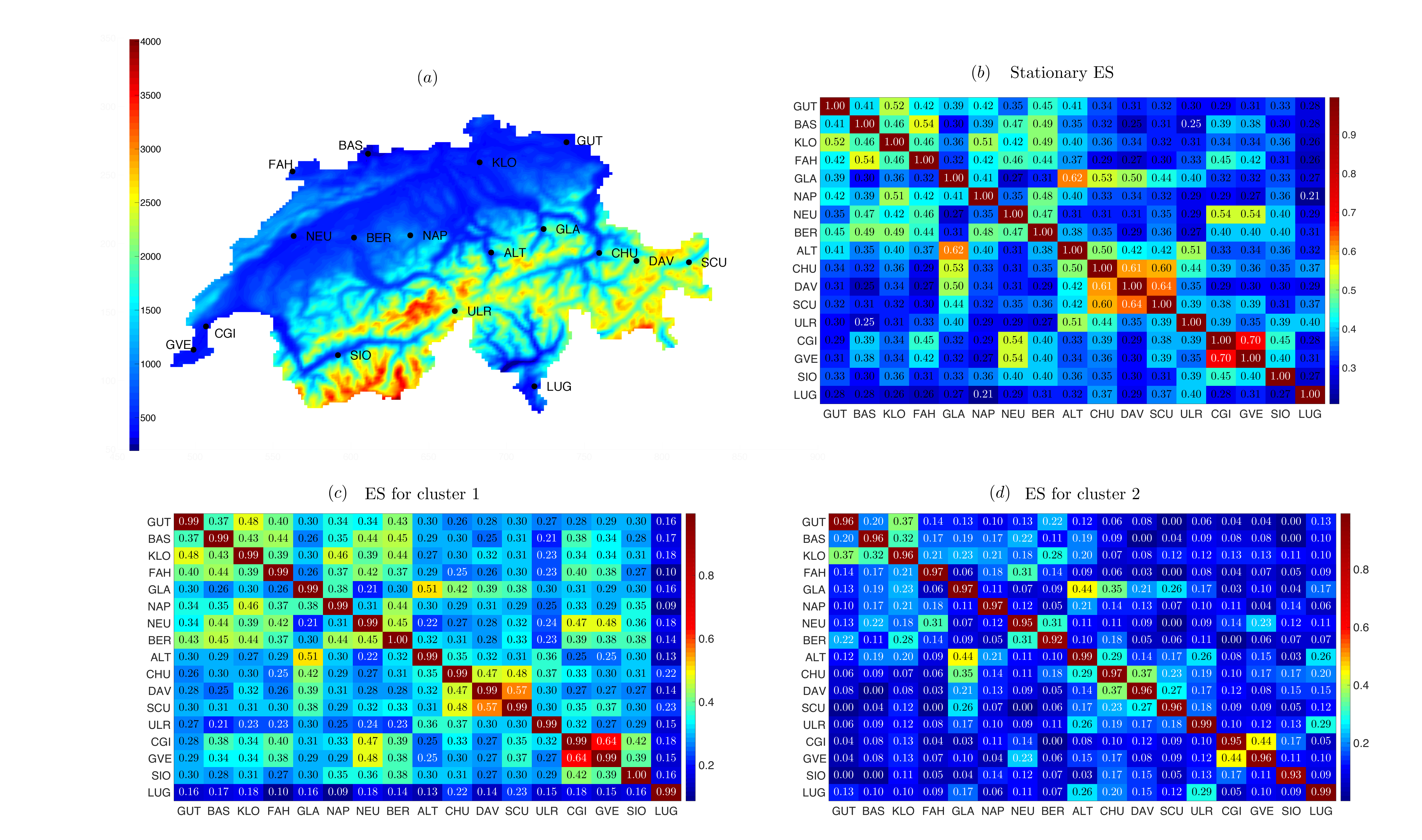}%
  \end{adjustbox}
\end{figure}

In the following the performance of the spatial FEM-BV-GPD is applied to daily accumulated precipitation data at 17 different locations in Switzerland from 1981-01-01 to 2013-01-01 (see Figure~\ref{fig:ES_cl12}, panel (a) for a map of the stations). %

At each location the threshold excesses were extracted with respect to a threshold defined by $QL + 0.00001$, where $QL$ is the $0.98$ quantile of the accumulated rainfall (for all, wet and dry, days) at this particular location. This threshold allows later for a bias correction of the shape parameter, as proposed in~\citep{WRCR012557, WRCR20707} for threshold excesses over the $0.98$ quantile threshold. The extraction of threshold excesses according to a 0.98 quantile would result on average in one additional extreme value for each location. %

The analysis involves a set of local covariates, measured at each location and a set of global covariates, being the same for each locations. The initial set of local covariates contains: (a) Wind (hourly maxima); (b) Temperature measured at 2 meters above the ground (hourly average); (c) Humidity, (d) Sun duration. The initial set of "global covariates", i.e., equal for all stations, which might influence the behavior of extreme precipitation events involves: (a) Total Solar Irradiance (TSI) averaged over one day~\citep{TSI,TSI2}; (b) North Atlantic Oscillation (NAO); (c) Arctic Oscillation (AO); (d) El Ni\~no Southern Oscillation (ENSO)~\citep{Trenberth}; (e) Periodical oscillation (Per), with $Per(t) = sin(\frac{2*\pi}{365}t)$; (f) Time-delayed ENSO, with a time lag of $3$, $12$ and $24$ month~\citep{rodo1997variations,knippertz2003decadal,ouachani2013power}, denoted in the following as $ENSO_3$ and $ENSO_{24}$. This set of covariates is reduced by considering only the uncorrelated covariates such that the regression analysis is carried out with respect to%
\begin{align}
	\label{eq:covasSpatialGPD}
	U_{s,t} = \lbrace Humid,\, TSI,\, Per,\, NAO,\, ENSO_3 \rbrace.
\end{align}
The covariates $U_{s,t}\in\mathbb R^{5}$ are scaled; the local ones to $[0,1]$ and the global ones to $[-1,1]$. %
So, the relative influences of covariates on trends in model parameters can be directly interpreted and compared. For regression analysis, the covariates are taken at the same time steps as the threshold excesses are observed. %

The spatial FEM-BV-GPD approach is configured as follows:  $K\in\lbrace 1,2,3 \rbrace$, $C_{T}\in\lbrace 10,20,\dots,100\rbrace$, in intervals of ten, and a set of different Lagrange multipliers is considered, denoted by $\lambda\in\lbrace 0, 0.0001, 0.001, 0.01, 0.1, 1\rbrace$. In order to obtain a global solution, FEM-BV-GPD is restarted with random initialization of $\Gamma(s,t)$ $600$ times. For the AO the maximal number of the subspace iterations is set to $1000$ and the convergency criterion is set to $1.0\times10^{-3}$. The optimal configuration is obtained for $ K=2$ , $C = 20$ and $\lambda = 0$ with $NLL_{\lambda} = 11887.528$ and $AICc = 24625.3443$. That is, the most optimal FEM-BV-GPD model describes the underlying behavior of threshold excesses over the 17 different location by two different locally stationary GPD-regression models and a nonstationary switching process $\Gamma(s,t)$ for each single location. The maximal number of model switches for each of the locations is $C=20$. In contrast, the stationary case, i.e., $K=1$, terminated with $NLL_{\lambda} = 12695.1097$ and $AICc = 25414.3046$. Thus, according to $AICc$, the optimal FEM-BV-GPD configuration does not overfit the data. %

\begin{figure}[htbp]
\vskip 0.2in
	\begin{center}
		\centerline{\includegraphics[scale=0.25]{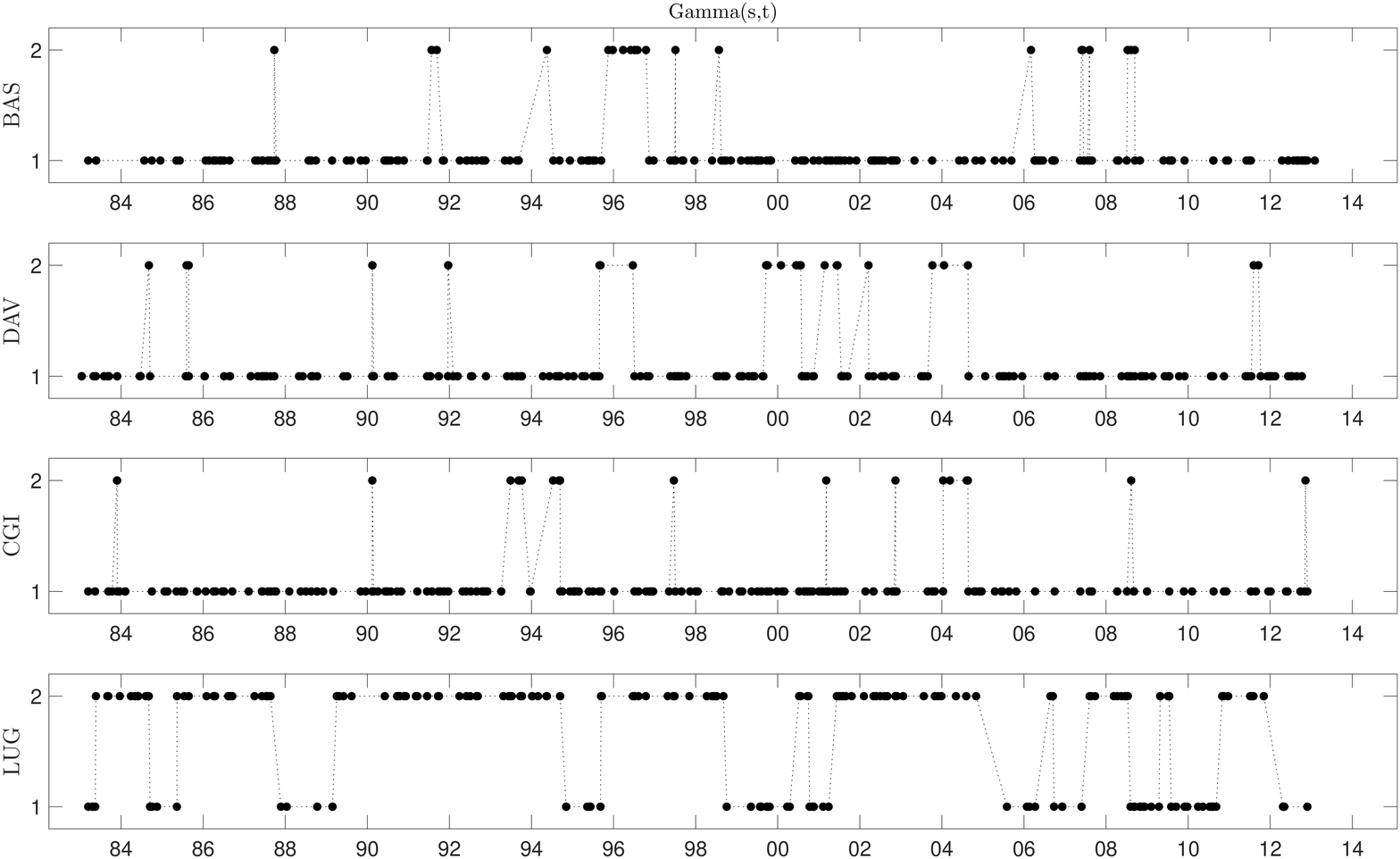}}%
		\caption{The realization of the optimal switching process for locations BAS, DAV, CGI, LUG. Each black dot indicates the timing (x-axes) and the affiliation (y-axes) to the first or the second cluster of an individual high precipitation event.}%
		\label{fig:spaFEMEVA_Gamma3}%
	\end{center}
\vskip -0.2in
\end{figure}

The optimal $\Gamma(s,t)$ is shown exemplifying for locations BAS, DAV, CGI, LUG in Figure~\ref{fig:spaFEMEVA_Gamma3} (representing North, East, West and South of Switzerland). %
The dominating cluster for all locations, except Lugano, is the first cluster. The switching process at location Lugano and thus also the behavior of threshold excesses is very different compared to remaining locations.  The models parameters for both clusters are summarized in Table~\ref{tab:Theta_confi}. %
%
\begin{table}[ht!]
\caption{The table contains the optimal FEM-BV-GPD parameters and its standard errors (corresponding to the rows indicated by $\pm$).}%
\label{tab:Theta_confi}
\vskip 0.15in
\begin{center}
\begin{tabular}{  c | c c  c  c  c c}
& \multicolumn{6}{ c }{ Relative influence of the covariates on GPD parameters}\\
 		   &off-set  & Temp.    & Humidity & TSI      & NAO      & ENSO$_3$     \\%
 \cline{1-7}
$\xi_1$    & -0.1049 &  -0.1280 &   0.0083 &  -0.0376 &   0.0156 &   0.0005 \\%
$\pm$      &  0.2820 &   0.3242 &   0.0484 &   0.0295 &   0.0750 &   0.0761\\%
$\xi_2$    & -0.4899 &   0.4920 &   0.1764 &  -0.0473 &  -0.0082 &  -0.0556 \\%
$\pm$      &  NaN    &   NaN    &   NaN    &   NaN    &   NaN    &   NaN\\%
$\sigma_1$ &  5.2598 &   4.9775 &   0.9028 &  -0.3796 &  -0.2372 &   0.5478\\%
$\pm$      &  3.2751 &   3.8279 &   0.6286 &   0.3713 &   1.0082 &   0.9214\\%
$\sigma_2$ & 30.8296 &   2.4295 &   2.6994 &  -0.1604 &  -0.0148 &  -0.2011 \\%
$\pm$      &  NaN    &   NaN    &   NaN    &   NaN    &   NaN    &   NaN\\%
\end{tabular}
\end{center}
\vskip -0.1in
\end{table}
For shape parameters the influence of temperature changes sign between cluster 1 and cluster 2, indicating a fatter tail with increasing temperature for cluster 2. For the scale parameter the local covariates temperature and humidity both contribute positively, the role of humidity compared to temperature is more important for cluster 2. For the scale and the shape parameters the relative influence of the local covariates is an order of magnitude stronger than that of the global covariates. We therefore do not discuss the global covariates in detail. %
The estimation of the model parameters for the first cluster is more robust, as it has a larger statistics, with in total 3000 data points. The parameter estimation in the second cluster has 679 data points. The confidence intervals were estimated using the Matlab command \verb|mlecov()|. It approximates the asymptotic covariance matrix of the maximum likelihood estimators of the parameters. In cluster 1 the confidence intervals for the shape parameter include the zero value: this could be an indicator for a constant shape parameter. And in fact, for a constant shape parameter the optimal FEM-BV-GPD model is obtained for $K=2,\,C=20$ with an $AICc = 24587.1269$ (i.e, according to $AICc$ this model is close to the fully nonstationary one). Also the dynamics described by the corresponding model parameters is very similar to the one obtained for the nonstationary shape parameter. In particular, the switching process is changing for many locations hardly. Thus, for further discussion we continue with the fully nonstationary case. %
The estimation of the confidence intervals failed for the second cluster because the estimated covariance matrix is not positive definite. However, as the statistics for the second cluster is smaller, we expect here larger confidence intervals compared to the first cluster. Additionally, following the results in~\citep{WRCR012557, WRCR20707}, a bias correction of the shape parameter, which accounts for the small statistics, could be implemented. We skip it for this analysis, as the correction appears to be more significant when the focus lies on predictive rather than on descriptive analysis.%

\begin{figure}[htbp]
	\vskip 0.2in
	\begin{center}
		\centerline{\includegraphics[scale=0.3]{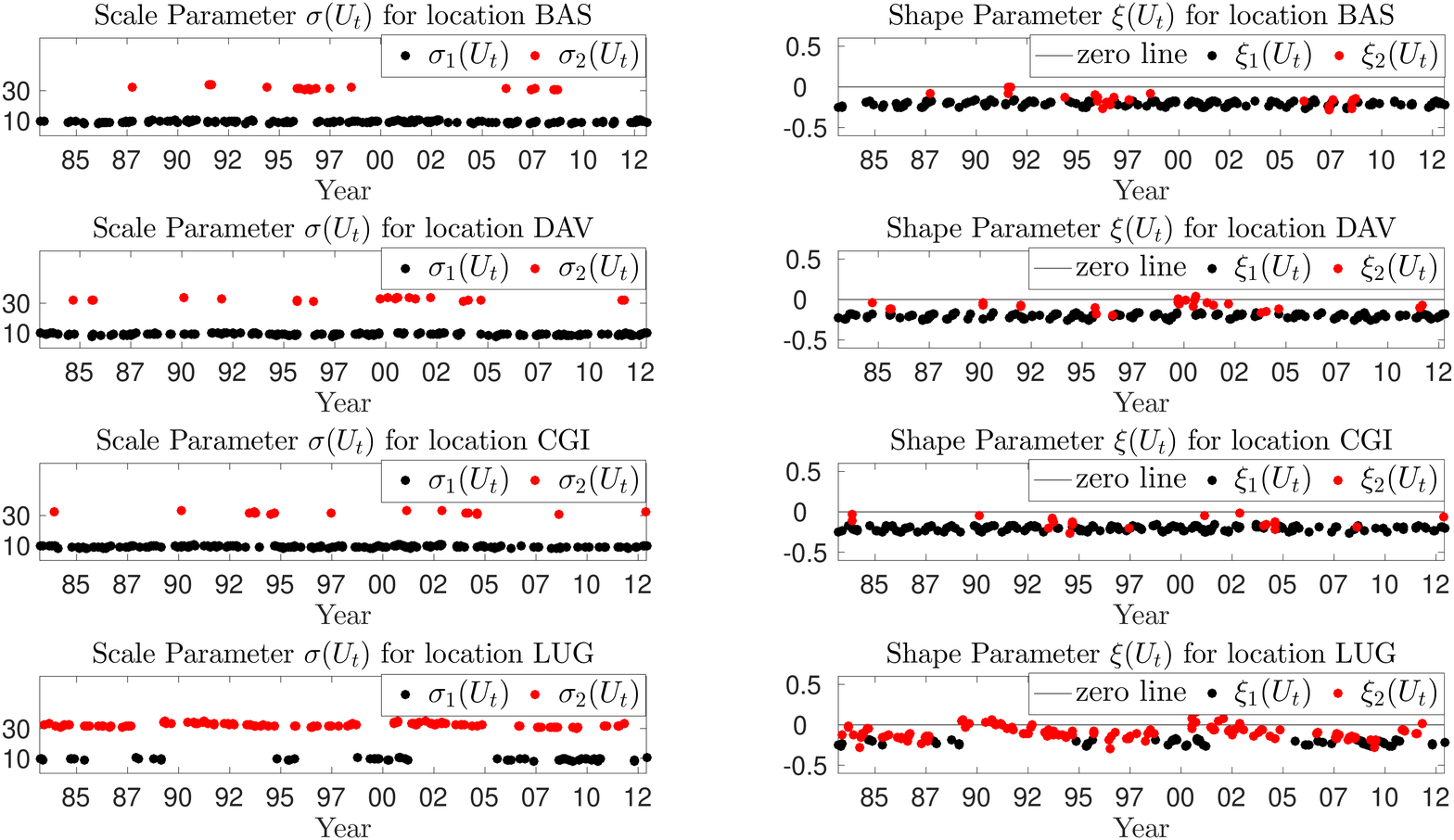}}%
		\caption{This figure shows the evaluation of the optimal FEM-BV-GPD model parameters for locations: BAS, DAV, CGI, and LUG. The right and left panels represent the scale and the shape parameters for each location, respectively. The black markers correspond to the first and the red to the second model.} 
		\label{fig:spaTheta_lugbernalt}
	\end{center}
\vskip -0.2in
\end{figure}

For the visualization issues the temporal behavior of the optimal parameters $\left(\xi_1,\sigma_1\right)$ and $\left(\xi_2,\sigma_2\right)$ can be evaluated by incorporating the corresponding switching process $\Gamma(s_j,t)$, $j=1,\dots,17$. To be consistent with the visualization of the switching process, the temporal behavior of the FEM-BV-GPD model parameters is shown for locations BAS, DAV, CGI, and LUG in Figure~\ref{fig:spaTheta_lugbernalt}. %
%
%
Threshold excesses for location Lugano are mostly affiliated to the second model, where the scale parameter exhibits larger values compared to the first model. Threshold excesses for the remaining locations exhibit an opposite behavior. The dominating behavior is represented by the first model. In total, cluster 2 describes the behavior of threshold excesses with higher intensity compared to cluster 1. %

Please note, we do not normalize and/or deseasonalize the accumulated rainfall, but cluster the threshold excesses according to their absolute intensity. For comparison, we also applied the spatial FEM-BV-GPD to normalized accumulated precipitation extremes (at each location the accumulated precipitation was normalized by the median of rainy days at this location). For this data, the optimal spatio-temporal FEM-BV-GPD model is also obtained for $K=2,\,C=20$. The underlying dynamics described by the optimal parameters $\Theta,\,\Gamma(s,t)$ differs from the one obtained for raw threshold excesses. To keep this section short, the results are not shown and are not discussed. 

In this context, it is important to emphasize, that in order to achieve the desired results when applying the spatial FEM-BV-GPD, the reader should adjust the data to the question he aims to address. For instance, our aim in this section is to investigate the cluster dynamics of the absolute intensity of threshold excesses. While the analysis of normalized threshold excesses addresses the relative intensity of threshold excesses. %

\begin{figure}[ht]
  \begin{adjustbox}{addcode={\begin{minipage}{\width}}{\caption{%
      This figure shows the QQ plot (exponential scale) for the application of FEM-BV-GPD. The dash-dot grey lines are the $95\%$ confidence bands for the QQ plots.}%
	   \label{fig:spaFEMEVA_qq}
  		\end{minipage}
		 },rotate=90,center}
      \includegraphics[scale=.35]{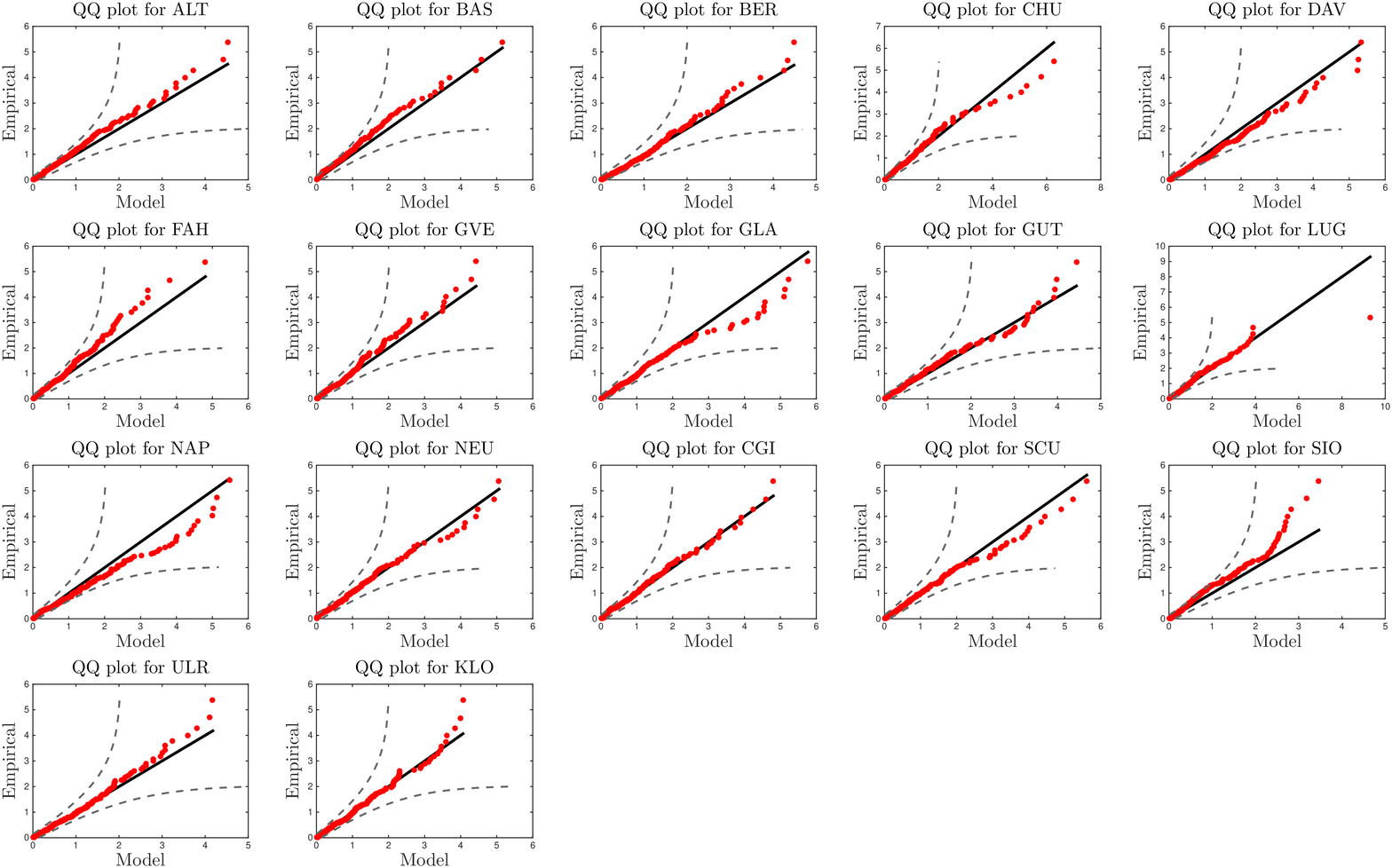}%
  \end{adjustbox}
\end{figure}

The model validation wrt. QQ-plots is shown in Figure~\ref{fig:spaFEMEVA_qq}. In order to account for the uncertainty in the QQ plots, confidence bands were estimated using the R function \verb|qqPlot()|. The results for all locations are within the confidence bands. %

\begin{figure}[ht]
      \includegraphics[scale=0.4]{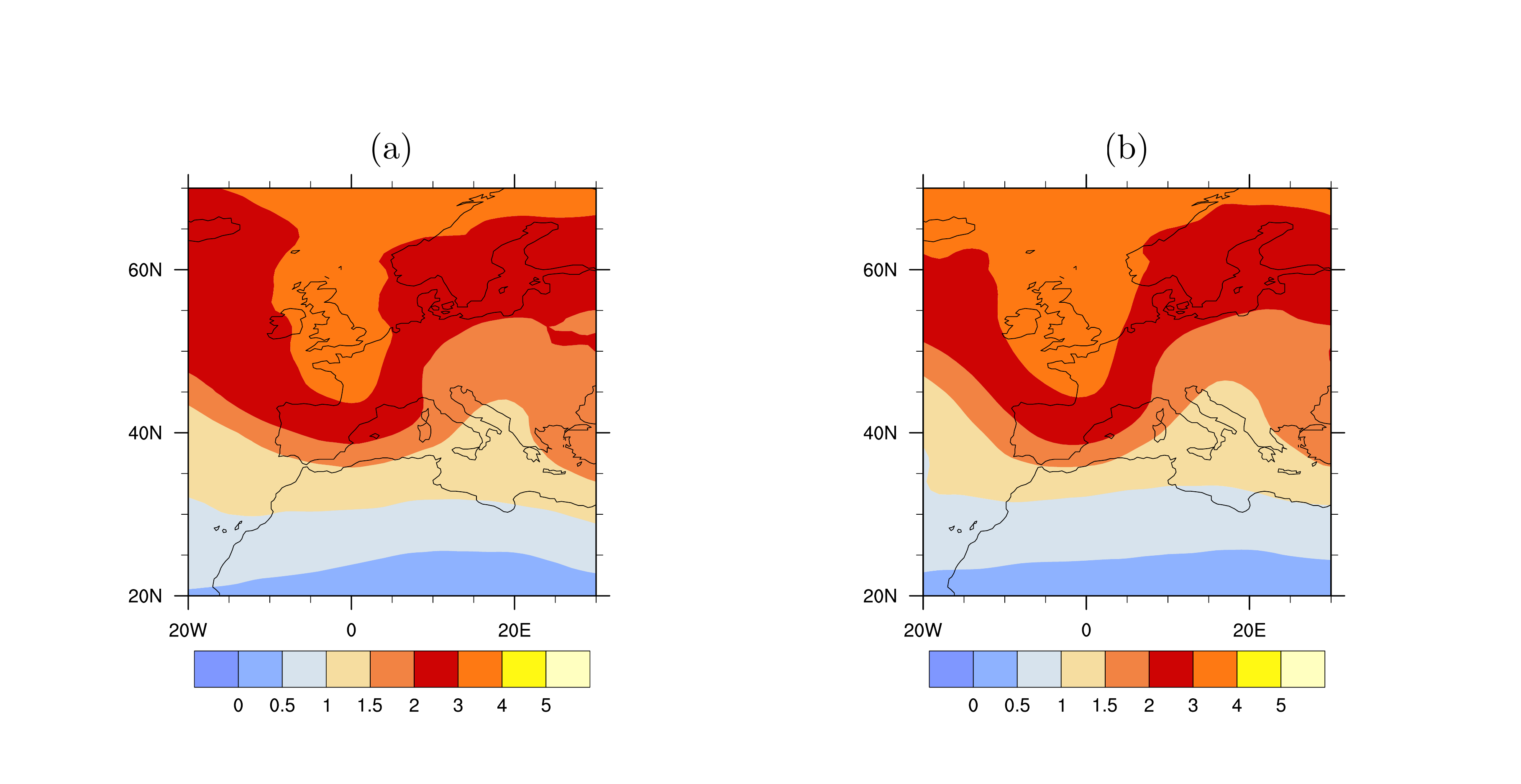}%
	  \caption{The panels (a) and (b) show the large-scale weather situation for the first and the second cluster. Shown is the composite mean upper-level vertically averaged PV (PVU, 1 PUV = 10-6 Km2/(kgs)). In both clusters the large-scale weather situation is similar with a trough over Western Europe and a ridge over Central and Eastern Europe.}%
  \label{fig:PVINT}
\end{figure}

In the next step, we investigate in more detail the meteorological formation of the switching process. The hypothesis is that the first cluster corresponds to large-scale and the second cluster to convective precipitation. To test this hypothesis we investigate the  characteristics of the atmospheric flow during the considered time period, i.e, from 1981-01-01 till 2013-01-01 during all cluster 1 and cluster 2 events in Lugano. We use the ERA-interim data set~\citep{dee2011era} to analyze the large-scale weather situation. We calculate composites of the vertically averaged (between 500hPa and 150hPa) Ertel Potential Vorticity to characterized the large-scale flow. We also calculate the integrated water vapor flux (IVT) following e.g.,~\citep{lavers2011winter}. To assess the stability we calculate composites of the most unstable Convective Available Potential Energy (CAPE). In Lugano both large-scale driven heavy precipitation and convective intense precipitation are linked to similar large-scale flow situations namely a trough over Western Europe and southerly to southwesterly flow at the surface~\citep{RN326,QJ2351}. Indeed the composite mean large-scale flow for all extreme precipitation events in Lugano in cluster 1 and in cluster 2, Figure~\ref{fig:PVINT}, is similar and shows a deep trough located over Western Europe. %

\begin{figure}[ht]
    \includegraphics[scale=0.4]{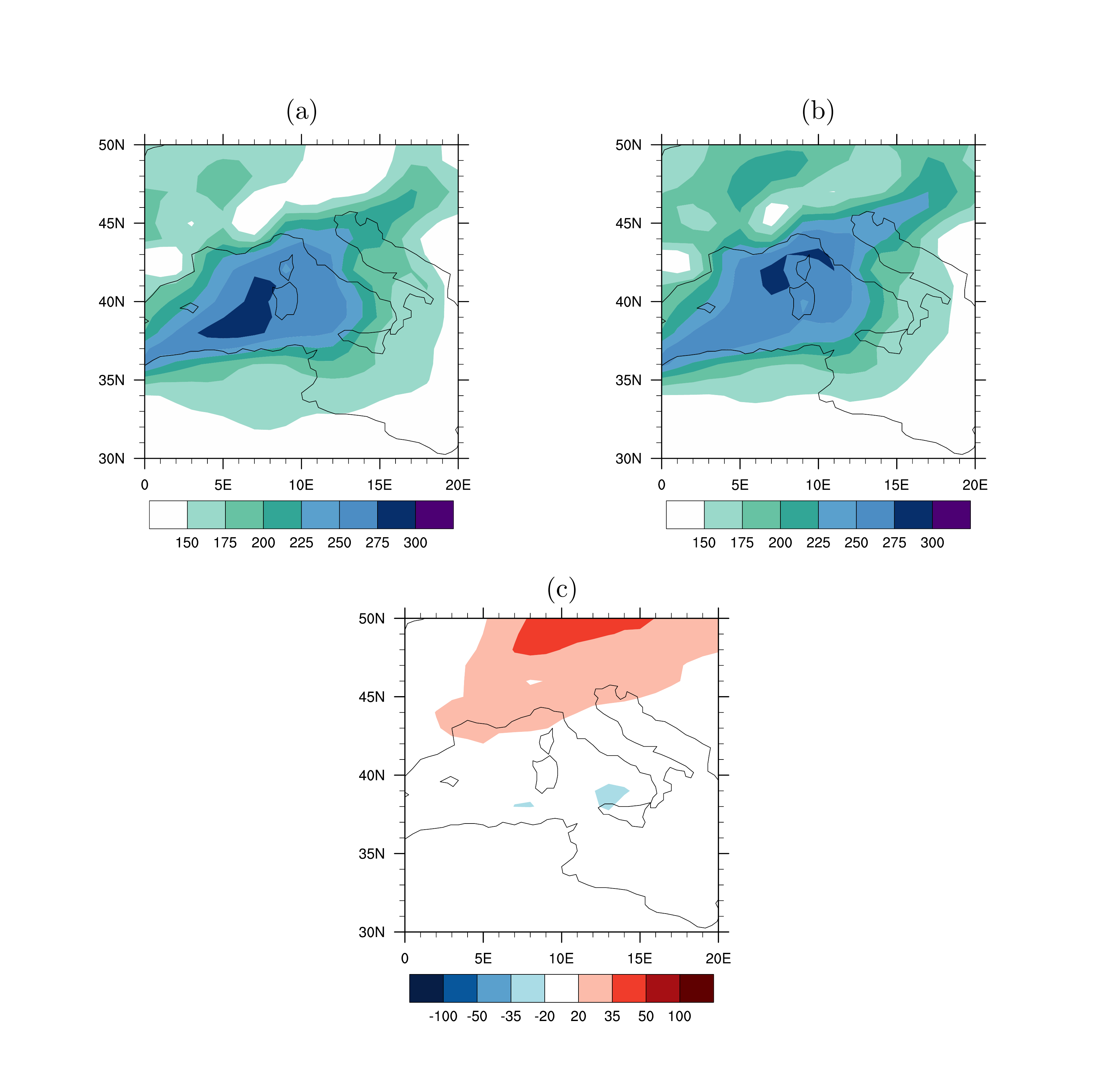}%
    \caption{The panels (a) and (b) show the composite mean vertically integrated humidity transport (kg/ms) for the first and the second cluster. Panel (c) shows the difference between the second and the first cluster. The moisture transport to the Alps is higher for the second cluster.}%
    \label{fig:IVT}
\end{figure}

\begin{figure}[ht]
  \includegraphics[scale=0.4]{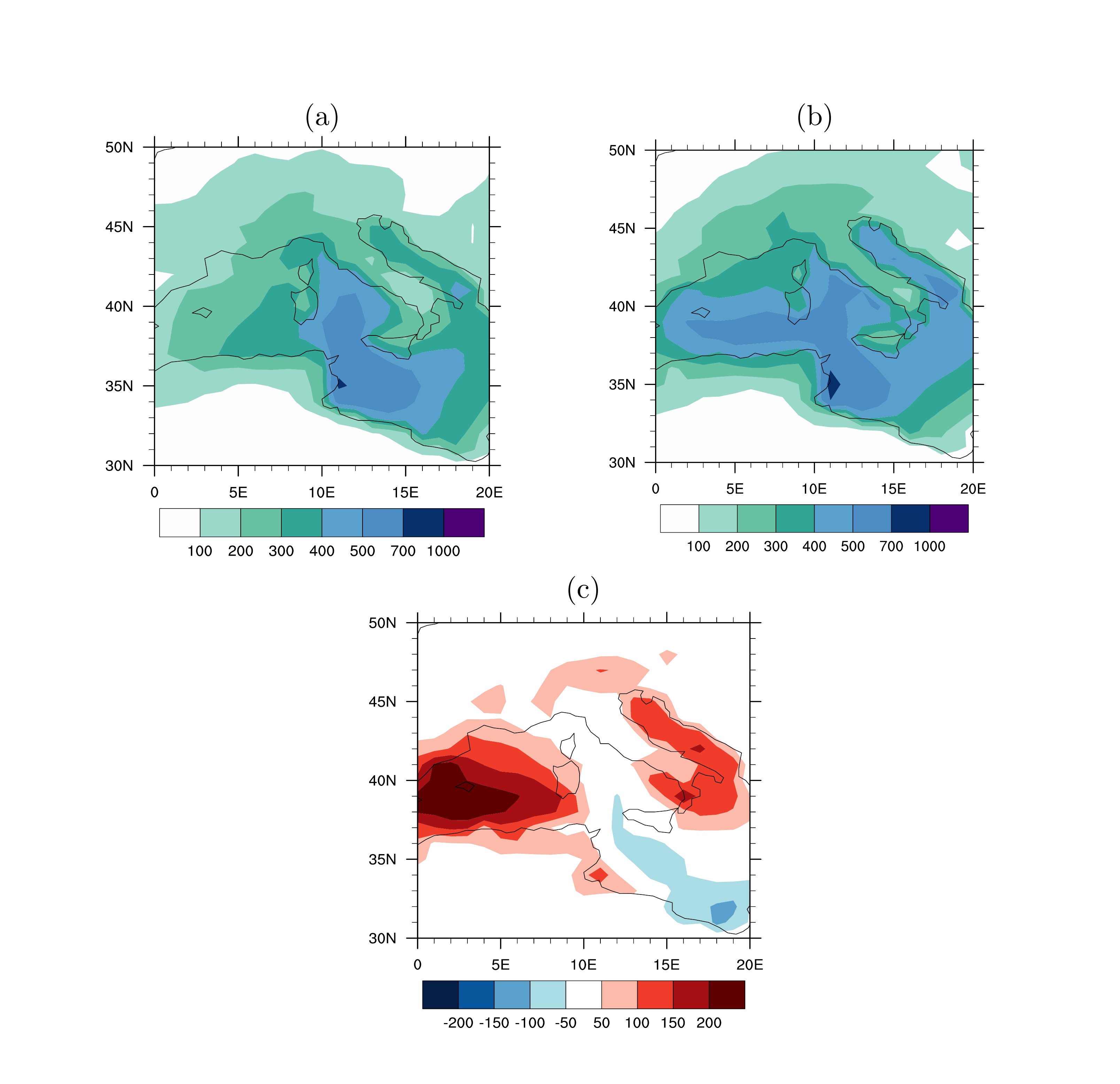}%
	\caption{The panels (a) and (b) show the composite mean CAPE (J/kg) for the first and the second cluster. Panel (c) shows the difference between the second and the first cluster. The atmospherical instability is higher over southern Switzerland for the second cluster.}%
  \label{fig:CAPE}
\end{figure}

The investigation of large-scale weather situation, Figure~\ref{fig:PVINT}, indicates similar behavior for both models/clusters: with a trough over Western Europe and a ridge over central and eastern Europe. The moisture transport to the Alps, Figure~\ref{fig:IVT}, is higher for the second cluster, matching the higher intensity of extremes in the second cluster. CAPE over the Ticino is higher for the second cluster, supporting the occurrence of more extreme precipitation resulting from (embedded) convection, Figure~\ref{fig:CAPE}. Lastly, in cluster 2 a larger fraction of days with lightening flashes within a radius of 3 km around Lugano is found (68 days of 142 with more than 5 lighting strokes in a 3km radius in cluster 2 and 8 out of 69 days with more than 5 lightning strokes in cluster 1). All of this points to the more convective nature of the events in cluster 2, however, the separation is not absolutely clear cut. Both clusters contain convective events. This is also in the nature of the precipitation, which can be primarily driven by the large-scale flow but also be locally enhanced by embedded convective cells.  %
%
%
Please note, for location Davos (DAV) some of extreme events in cluster 2 occur in winter, these are clearly not convective and this would be an argument against our hypothesis.%

It can be concluded that the occurrence of extremes in the second cluster (the more extreme extremes) was related to higher instability and a higher humidity transport towards the Alps. Thus, further analysis for more locations are required in order to test the hypothesis that the two obtained clusters correspond to convective and large-scale precipitation. This remains for future work. %

Further, as already discussed, by identifying locations that exhibit similar dynamics of the switching process we can get insights into the underlying spatial dependence structure. For this purpose we make use of the event synchronization measure (ES)~\citep{malik2012analysis} which estimates the nonlinear correlation among locations. Hereby, we will estimate the stationary ES by investigating the occurrence of all events and the cluster-wise ES by investigating the occurrence of events for each cluster. The ES matrices are presented in Figure~\ref{fig:ES_cl12}.%
%
%
%
%
Complete similarity is ensured when the ES entry reaches the value $1$. In the following, we consider two locations as correlated when the corresponding ES value is larger than $0.35$. %
The stationary ES exhibits larger correlated regions in adjacent geographic areas, for instance, (a) GUT, BAS, KLO, FAH, (b) ALT, CHU, DAV, SCU, URL, and (c) URL, CGI, GVE, SIO. In the FEM-BV-GPD context, the above correlated regions remain mainly for cluster 1, while cluster 2 retains only the strong correlation, e.g., (a) KLO, GUT, (b) ALT, GLA, and (c) GVE, CGI. Further, according to the stationary ES, location Lugano, which represents the regions south of alps, exhibits weak correlation to locations CHU and SCU. This indicates a connection between the East and the South regions of Switzerland and disagrees with results obtained in ~\citep{schmidli2005trends} and ~\citep{dailyExtremePrecipitationMeteoSwiss97}. %
In contrast to the cluster-wise ES: For both clusters location Lugano is not correlated with the rest of Switzerland. Such that the cluster-wise ES points to the four major climatic regions for precipitation in Switzerland: North, East, South, and West~\citep{schmidli2005trends}, and they are partially in agreement with the scientific report of MeteoSwiss, where the seasonal variability of extreme precipitation was analyzed~\citep{dailyExtremePrecipitationMeteoSwiss97}. However, verification of this adjacent regions for extreme precipitation requires further analysis with more locations. %
Moreover, the cluster-wise investigation of ES provides an additional argument for the hypothesis that the two obtained clusters correspond to convective and large scale precipitation: The generally higher correlations in cluster 1 and lower values in cluster 2 point to more convective events in cluster 2. %
%
%
%
%
%
%
\section{Conclusion} 
\label{sec:conclusion}
Based on the Generalized Pareto Distribution (GPD) and a nonparametric Finite Element time-series analysis Methodology (FEM) with Bounded Variation of the model parameters (BV), we presented a nonstationary and nonhomogenous framework for regression analysis of spatio-temporal threshold excesses in a presence of systematically missing covariates. We showed that under weak assumptions the influence coming from systematically missing covariates can be reflected by a nonstationary and nonhomogenous offset. %
FEM-BV-GPD resolves the spatio-temporal behavior of the model parameters by a set of local sparse stationary GPD models and a nonparametric, persistent hidden switching process. The set of GPD models is accessible for all locations, while each location has its own switching process. The affiliation of extremes with respect to the resulting hidden switching process enables their spatio-temporal regression based clustering - without a necessity to impose a strong assumption of independence assumption between different locations. Instead, we assume the locations conditionally-dependent on the state of the latent process $\Gamma(s,t)$ that we infer during the parameter identification procedure. %
 
The resulting spatial FEM-BV-GPD can be employed as a robust exploratory sparse regression analysis tool for spatio-temporal extremes, that enables to identify the significant resolved covariates and to account for the influence from the systematically missing ones. The right choice of covariates allows to understand the future trends in the behavior of extremes, e.g.,~\citep{tramblay2012extreme}, and should not be negligible. %
FEM-BV-GPD can be used to describe the behavior of return levels and periods in the past in such a manner that possible trends become visible. However, it is not directly applicable for predictions of the return levels in the future. This task is hampered by the fact that the behavior of extremes is described by a set of local model parameters and a nonparametric, nonstationary switching process. This means that there is no closed formulation for the underlying dynamics of $\Gamma(s,t)$ which is required for prediction. To approach this problem, one can either try to find an extended set of covariates in order to resolve the observed dynamics, see e.g.,~\citep{Horenko2}, such that the optimal model is obtained for $K=1$ or to find a model for the spatio-temporal process $\Gamma(s,t)$ like the logistic regression or the more complex neural networks based approaches. %
A-posteriori, $\Gamma(s,t)$ provides a pragmatic description of the corresponding spatial dependence structure by grouping together all locations that exhibit similar behavior of the switching process. %

The proposed methodology was realized as a gradient-free MCMC based optimization techniques - combined with numerical solvers for constrained, large, structured linear problems. Demonstration of spatial FEM-BV-GPD was performed on threshold excesses of daily accumulated precipitation over 17 different locations in Switzerland. The optimal FEM-BV-GPD description of the extremes was obtained for two local clusters/models and a nonstationary switching process $\Gamma(s,t)$ for each of the locations. It is demonstrated that this optimal model implies that the switching process for location Lugano exhibits a completely different behavior compared to all the other locations. %
We investigated the behavior of the switching process and revealed that in southern Switzerland (Lugano) the occurrence of extremes in the second cluster, responsible for the more extreme extremes, is related to higher atmospheric instability and the higher humidity transport towards the Alps. Additionally, we could find some indications for the hypothesis that the two obtained clusters of the extreme precipitation events correspond to large-scale (first cluster) and convective precipitation (second cluster). Further, by applying the Event Synchronization (ES) measure we obtained a pragmatic description of the underlying dependency structure. The result suggests four major statistically-significant extreme climatological regions in Switzerland: North, East, South, and West. However, for verification of these results further analysis with more locations and measurements will be necessary. This remains for future work.%
%
%
%
%
%
\section{Acknowledgments} 
\label{sec:acknowledgments}
The precipitation data for this paper was retrieved from MeteoSwiss (\url{www.meteoswiss.admin.ch}). The covariate TSI was retrieved from Physikalisch Metorologisches Observatorium Davos/World Radiation Center (\url{http://www.pmodwrc.ch/pmod.php?topic=tsi/composite/SolarConstant}). The covariates NAO and ENSO were retrieved from National Oceanic and Atmospheric Administration (\url{ftp://ftp.cpc.ncep.noaa.gov/cwlinks/}). We thank Simona Trefalt for the lightning data. I. Horenko was partly funded by the Swiss Platform for Advanced Scientific Computing, Swiss National Research Foundation Grant 156398 MS-GWaves, and the German Research Foundation (Mercator Fellowship in the Collaborative Research Center 1114 Scaling Cascades in Complex Systems). During this work O. Kaiser was supported by the Swiss Platform for Advanced Scientific Computing, Swiss National Research Foundation Grant 156398 MS-GWaves. %
\bibliographystyle{plainnat}
\bibliography{extr_ev_multdim_nonstat,eva,climateDynamics,tsAna,statistics}

\end{document}